# A Single Cluster Covering for Dodecagonal Quasiperiodic Structure

Longguang Liao, Zexian Cao[*]

Institute of Physics, Chinese Academy of Sciences, P. O. Box 603, Beijing 100190, China

**Abstract**

Single cluster covering approach provides a plausible mechanism for the formation and stability of octagonal and decagonal quasiperiodic structures. For dodecagonal quasiperiodic pattern such a single cluster covering scheme is still unavailable. Here we demonstrated that the ship tiling, one of the dodecagonal quasiperioidic structures, can be constructed from one single prototile with matching rules. A deflation procedure is devised by assigning proper orientations to the tiles present in the ship tiling including regular triangle, 30°-rhombus and square, and fourteen types of vertical configurations have been identified in the deflated pattern, which fulfill the closure condition under deflation and all result in a T-cluster centered at vertex. This result can facilitate the study of physical properties of dodecagonal quasicrystals.

## I. Introduction

Quasicrystals are solids with long-range positional order and orientational order, but non-periodic translational order[1,2]. In the quasicrystals, 5-fold, 8-fold[3], 10-fold[4] and 12-fold[5] rotational symmetries may appear, which are forbidden in the ordinary crystals. Unlike periodic crystals which can be constructed from one simple unit cell, an ideal quasicrystal is often modeled as a tiling with two or more distinct building blocks. In the 2D case, the most renowned model is the Penrose tiling[6] which comprises of fat and thin rhombuses subject to particular matching rules that constrain the ways how neighboring rhombuses join together edge-to-edge, coercing them into a fivefold-symmetric pattern.

---

[*] Corresponding author. Tel.: 86-10-82649136; E-mail: zxcao@iphy.ac.cn

The Penrose tiling and alikes are a powerful, primordial model for understanding the structure of quasiperiodic patterns, but unfavorable to explain the formation and stability of quasicrystals since the rigid matching rules are too strict in comparison to the real situation of crystal growth. Theoretical doubts about its validity have remained[7]. Alternatively, the cluster covering model, which is a new approach and can provide a plausible mechanism for the formation and stability of qusicrystals, has been raised and welcomed[8]. The cluster covering model demands only a single repeating unit, which is similar to the concept of unit cell in crystals. However, in the cluster covering model the repeating clusters, when subjected to certain rules to make a quasiperiodic pattern, often have to be allowed to partially overlap. Gummelt once proposed a decagonal covering scheme and showed that a quasiperiodic structure equivalent to the Penrose tiling could be constructed from decorated decagons[8]. Jeong and Steinhardt demonstrated that by maximizing the density of a chosen cluster comprising of several fat and thin rhombuses it can give rise to the Penrose tiling[9].

Many other works have been devoted to the single cluster covering model of quasicrystals in the past years, and various single cluster covering models for decagonal and octagonal two-dimensional quasicrystals have been established[10,11]. It is now still a challenge to extend this approach to the dodecagonal quasiperiodic structure. Ben-Abraham and his coworkers, after proposing several "almost covering" patches, claimed conclusively in 2001 that "a complete covering of dodecagonal quasiperiodic structures requires two clusters"[12]. In a recent paper[13], one of the current authors, in studying the deflation of Stampfli- Gähler tiling, noticed the existence of a Turtle-like cluster tile, the T-cluster, and at that time it came to us that this type of special cluster tile may make a perfect covering of the ship tiling, one of the dodecagonal quasiperiodic structures[12].

In the current paper, we will show that the ship tiling, or the Stampfli-Gähler tiling[14] (see Fig.1 therein), of the dodecagonal quasiperiodic structures, can be perfectly covered by using only T-clusters. It all begins with the deflation procedure[15] for the ship tiling, through which a perfect ship tiling can be generated starting from any of the three tiles: square, regular triangle and 30°-rhombus. The matching rules will be explained, and the enclosure relation in the fourteen vertical configurations resulting from the deflation of the ship tiling will be demonstrated, which is the key point for the proof of the single cluster covering scheme, as the deflation of all the fourteen vertical configurations gives rise to a T-cluster centered at the vertex.

## II. Proof Details

### A. Deflation rules for ship tiling

Several types of dodecagonal quasiperiodic structures have been proposed. In 1986, Stampfli constructed a dodecagonal quasiperiodic pattern from three distinct tiles: a square, a regular triangle, and a 30°-rhombus[16]. By using cut-and-projection method Gähler pushed this idea one step forward and obtained a structure similar to that of Stampfli, but the number of the rhombuses used is enormously reduced[17]. Following the nomenclature of Ben-Abraham and his coworkers, this pattern was dubbed the ship tiling[12]. On the other hand, Socolar presented a dodecagonal pattern comprising of squares, regular hexagons, and 30°-rhombi, which is now called the Socolar tiling[15]. Ship tiling and the Socolar tiling are locally derivable from each other, or they are locally equivalent[18]. Here we just consider the ship tiling.

Like other quasiperiodic structures, the ship tiling is a self-similar pattern. This can be verified by applying the deflation procedure[15], a kind of self-similarity transformation, to the ship tiling. The deflation procedure we devised, as illustrated in Fig.1, replaces a square by five squares, twenty regular triangles and four 30°-rhombuses (of the next generation), a regular triangle by three squares and ten regular triangles, and a 30°-rhombus by two squares, twelve regular triangles and three 30°-rhombuses. The tiles of the second generation are scaled by a factor of $\alpha = 2 - \sqrt{3}$ with regard to the first generation. Remarkably, in so doing all the tiles, the square, the regular triangle and the 30°-rhombus, must be labeled with a specific orientation. To indicate the orientation of the deflation result, tiles have been marked with arrows (Fig.1). The specification of the orientation for the resulting tiles is an essential element in the deflation procedure, it helps coordinate the deflation of edge-sharing units following the deflation rules. If the arrows are ignored, there are then certain tiles generated through the given deflation rules that may break the demanded 12-fold symmetry. The deflation configuration for a square, as shown in Fig.1a, retains the mirror symmetry about the vertical bisection line, the original D4 symmetry of the square has been, however, broken. For this reason, an arrow along the vertical bisection line is attached in the square of the first generation to indicate the orientation of deflation result which possesses only mirror symmetry with regard to this arrow. For the deflation of a regular triangle, as shown in Fig.1b, the full $D_3$ symmetry of the original triangle is preserved. The arrow to specify the orientation of the deflation configuration can thus point to any vertex of the original regular triangle along the bisection line through it, which coincides with the orientation of one of the three second-generation squares. Hence for clarity, here only three small arrows are given in Fig.1b, which all can indicate the orientation of the triangle of the first generation, and at the same time each of them also specifies the orientation of the square of second generation it resides in. In the case of a rhombus the situation is somewhat more complicated. As shown in Fig. 1c, although the two squares and the two rhombuses at the far end of the wings, with orientation also considered, are still mirror-symmetrical, the deflation configuration as a whole is however not mirror-symmetrical as the central rhombus, perpendicular to the parent rhombus, has to take a preferred orientation. Thus the orientation of this first-generation rhombus is indicated by a bent arrow along the short diagonal where the arrow head defines the preferred orientation of the central rhombus of the second generation. In brief, Fig.1 exemplifies the deflation scheme here concerned for the ship covering of the dodecagonal quasicrystal structure. The readers are reminded that arrows marked on tiles are introduced to indicate the orientation of the deflation in the deflation rules. It is an aid for the generation of the ship tiling via deflation rules. Once the ship tiling has been generated, those arrows can be removed, or overlooked.

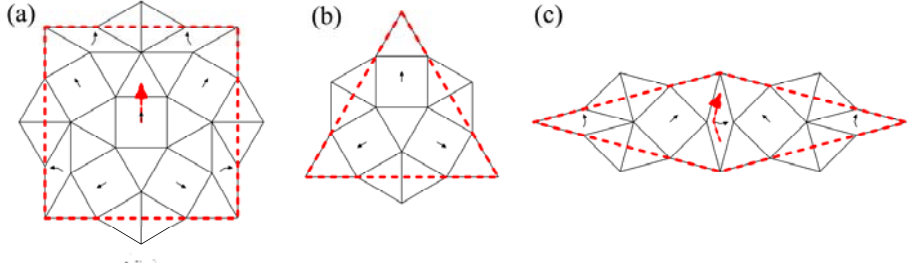

**Fig. 1** (color online)  Deflation rule for the ship tiling, as illustrated by the corresponding result (thin, black solid lines) of its application to the three original tiles (bold, red dashed lines): (a) a square; (b) an equilateral triangle; and (c) a 30°-rhombus. The arrows in the original tiles indicate the orientation of the resulting configurations, generally of a reduced symmetry. For the equilateral triangle, it has three equivalent directions which are, respectively, the direction of the three squares in the resulting configuration.

Using the deflation-inflation method, we can generate the dodecagonal ship tiling starting from either a square, a regular triangle, or a 30°-rhombus. The first step of the transformation is deflation as mentioned above, which changes a chosen tile into several smaller tiles of various types, scaled by a factor of $2-\sqrt{3}$ with regard to the corresponding original ones. The second step is inflation, which rescales the new pattern by a factor $2+\sqrt{3}$. By repeating this procedure again and again, we can get a patch of dodecagonal ship tiling of an arbitrary size.

**B. Vertex types in ship tiling and enclosure condition**

The ship tiling has a notable characteristic in edge sharing, that is, a square always joins edge-to-edge with four regular triangles, and so does a 30°-rhombus, while a regular triangle share its edges either with two squares plus one regular triangle, or two regular triangles plus one square, or one regular triangle, one square plus one rhombus. This characteristic of edge sharing conforms with the aforementioned deflation rules, but we shall not go into the proof of this point here.

On the basis of the deflation rules and the characteristic edge sharing, we find that if the arrowed directions are also taken into account, there will be fourteen possible vertex types in the ship tiling, which are listed in Fig. 2. To facilitate the description below, we name each vertex type with the letter V followed by two numbers. For example, V5-3 refers to the third type of a vertex joining five tiles. The first eleven vertex types in Fig.2 can be easily identified by applying the deflation rules. To find out the last three vertical configurations, i.e. V7-2, V7-3 and V7-4, one can apply the deflation rules to the first eleven vertical configurations. When V4-1 and V4-3 are concerned, the deflation procedure results in V7-4 and V7-3, respectively. When the other nine configurations are concerned, the vertical configuration V7-2 can be obtained. From the results of applying the deflation procedure to V7-2, V7-3 and V7-4, we can also find V7-2. This is to say that for the fourteen vertex types in Fig.2, the closure condition is fulfilled.

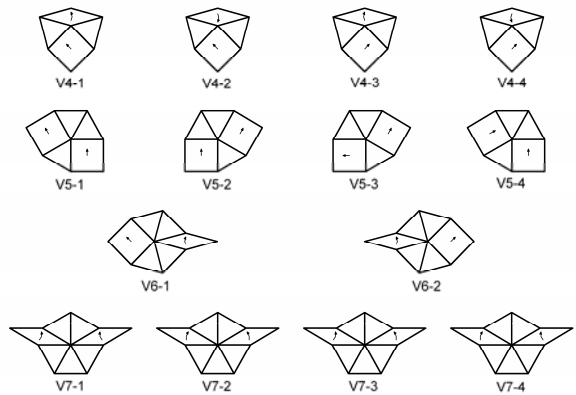

**Fig. 2**  The fourteen possible vertical configurations present in the ship tiling, with the orientation for further deflation of the tiles indicated by arrows.

Also a feature of the arrow directions for the two rhombuses in V7-1, V7-2, V7-3

and V7-4 is noticeable. The arrows for the two 30°-rhombuses in those four configurations always point to the same side of the straight line joining the bottom edges of the two 30°-rhombuses. As each 30°-rhombus has two possible orientations, the vertical configurations V7-1, V7-2, V7-3 and V7-4 are in fact characterized by the combination of the orientations of the two arrows. Just for this useful feature, once a 30°-rhombus and the vertex to be shared are fixed, the position of the other 30°-rhombus will be determined, which in turn can define the position and orientation of the corresponding T-cluster. Details will be given below.

**C. Complete covering of ship tiling by a single cluster**

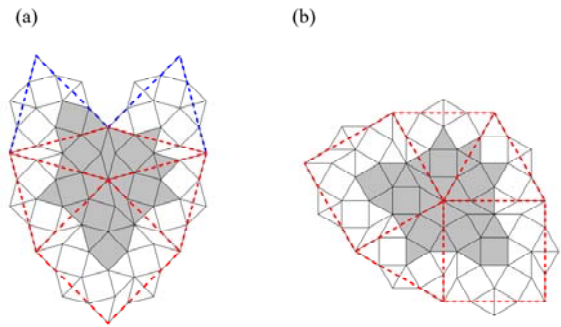

**Fig. 3** Deflation result (thin, black solid lines) of two typical vertical configurations (bold, red dashed lines): (a) V4 augmented with two triangles; (b) V5. For all the vertical configurations in Fig.2, the deflation gives rise to a T-cluster (in dark gray) about the original vertex.

Now we apply the deflation procedure to the aforementioned fourteen vertical configurations, from the results we can find a striking fact that each vertex of the ship tiling of first generation corresponds to a special cluster tile in the second-generation structure. Two cases are illustrated in Fig.3, for instance, with the shape of this special cluster tile shaded in the figure. This special cluster tile comprises of seven squares, twenty regular triangles and two 30°-rhombuses. Since its shape looks like a turtle, it is referred to as a T-cluster in our previous publication[14]. It is worth mentioning that T-clusters appearing in the results of deflation on V4-1, V4-2, V4-3 and V4-4 are incomplete. Yet, considering the characteristics of edge sharing, if two supplementary regular triangles are added to them, as is always reasonable, then a perfect T-cluster can be found for all occasions, see Fig. 3a.

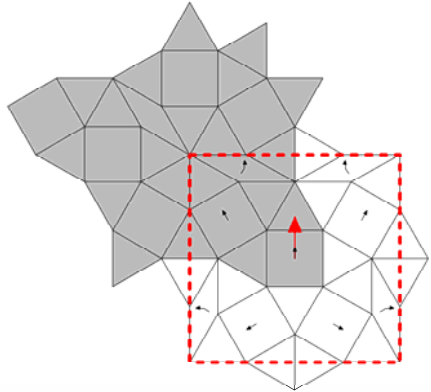

**Fig. 4**   A deflation-resulted T-cluster (thin, black solid lines) centered at one vertex of the original square (bold, red dashed lines). The four 30°-rhombuses with the attached square in the original square, when arrowed as in the figure, i.e., mirror-symmetrical about the arrow of the original square, result by deflation in four T-clusters centered at each vertex, so oriented that they completely cover the original square.

Now let us discuss the possibility of covering the ship tiling solely by the T-cluster. First, if three T-clusters in the deflation configuration are centered at the three vertices of the parent regular triangle, obviously these three T-clusters completely cover the parent regular triangle. This is also true for the case of a 30°-rhombus, if the four T-clusters are centered at the four vertices of the parent 30°-rhombus. These two situations are quite simple, thus not sketched here. In the case of a square, it is somehow a little complicated. There are four offspring 30°-rhombuses within the deflation configuration for a square, and each of the four offspring 30°-rhombuses has a well-specified orientation, see Fig.1a. Combining with the aforementioned fact that each vertex of the first generation tiles corresponds to a T-cluster of the second generation centered at the vertex, we can easily construct the four T-clusters correspond to each of the four vertices of a square. Fig. 4 shows an offspring T-cluster centered at the upper-left vertex of a square. And it is easy to verify that the four offspring T-clusters corresponding to the four vertices of a square can make a perfect covering of the square.

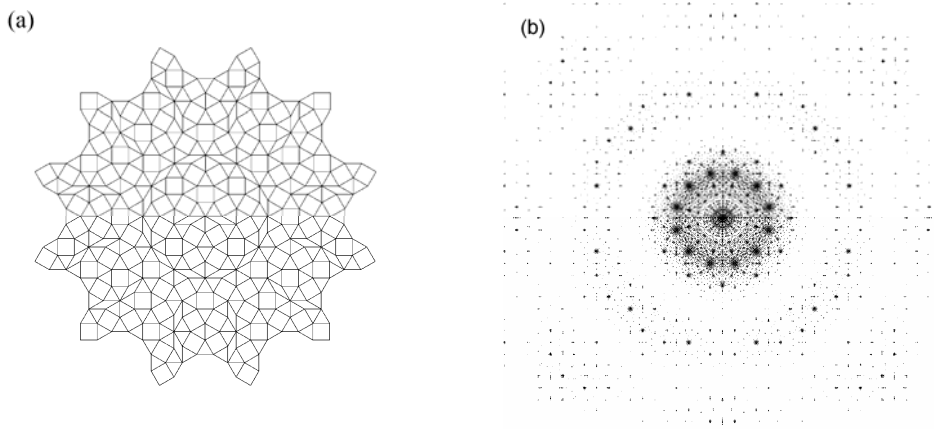

Fig.5 A finite size covering with 20 T-clusters, and the Fourier transformation of a finite size covering with 308 T-clusters showing distinct 12-fold symmetry.

So far, we see that a regular triangle, as well as a square and a 30°-rhombus, can be completely covered by three or four offspring T-clusters, the shaded region in Figs.3-4, centered at the corresponding vertices. Since the ship tiling is constructed from regular triangles, 30°-rhombuses and squares, and that each vertex of the first generation tiles corresponds to an offspring T-cluster, we can therefore safely draw the conclusion that the ship tiling can be perfectly covered uniquely by the T-clusters. Fig.5a displays the covering of a finite area by using 20 T-clusters, and the Fourier transformation of a patch covered by 308 such T-clusters reveals a typical 12-fold diffraction pattern.

**III. Summary**

In summary, we demonstrated for the first time the deflation procedure for the ship tiling of dodecagonal quasicrystal structure using properly oriented tiles: regular triangle, 30°-rhombus and square. In this process, fourteen possible types of vertical configurations appeared in the deflation result of ship tiling have been identified based on the deflation rules and the characteristic edge-sharing among the tiles. Through applying the deflation procedure to the fourteen vertical configurations, it is shown that ship tiling as one of the dodecagonal quasicrystal structures can be perfectly covered by one single tile—the T-cluster. This work is a step forward in constructing the covering for dodecagonal quasiperiodic structures in comparison to the previous 'almost covering' scheme, which may facilitate the study of physical properties of the dodecagonal quasicrystals.

**Acknowledgments**

This work was financially supported by the National Natural Science Foundation of China Grant nos.10974227, 51172272, and 10904165, and by the National Basic Research Program of China grant nos. 2009CB930801 and 2012CB933002, and by the Innovation Program of the Chinese Academy of Sciences.